\documentclass[12pt,preprint]{aastex}

\newcommand{\pad} {\partial}
\newcommand\simlt{\lower.5ex\hbox{$\; \buildrel < \over \sim \;$}}

\slugcomment{ApJ, to be submitted}

\shorttitle{Formation of hard VHE blazar $\gamma$-ray spectra}
\shortauthors{E. Lefa et al.}

\begin{document}

\title{Formation of hard very-high energy spectra of blazars in leptonic models}

\author{E. Lefa\altaffilmark{1,2}}
\affil{$^1$ Max-Planck-Institut f\"ur Kernphysik, P.O. Box 103980, 69029 Heidelberg, Germany}
\affil{$^2$ Landessternwarte, K\"onigstuhl 12, 69117 Heidelberg, Germany}
\email{eva.lefa@mpi-hd.mpg.de}
\author{F.M. Rieger\altaffilmark{1}}
\affil{$^1$ Max-Planck-Institut f\"ur Kernphysik, P.O. Box 103980, 69029 Heidelberg, Germany}
\and
\author{F. Aharonian\altaffilmark{1,3}}
\affil{$^1$ Max-Planck-Institut f\"ur Kernphysik, P.O. Box 103980, 69029 Heidelberg, Germany}
\affil{$^3$ Dublin Institute for Advanced Studies, 31 Fitzwilliam Place, Dublin 2, Ireland}

\begin{abstract}
The very high energy (VHE) $\gamma$-ray spectra of some TeV Blazars, after being corrected for
absorption in the extragalactic background light (EBL), appear unusually hard, which poses
challenges to conventional acceleration and emission models. We investigate the parameter
space that allows the production of such hard TeV spectra within time-dependent leptonic
models, both for synchrotron self-Compton (SSC) and external Compton (EC) scenarios.
In the context of interpretation of very hard $\gamma$-ray spectra, time-dependent considerations
become crucial because even extremely hard, initial electron distributions can be significantly
deformed due to radiative energy losses.
We show that very steep VHE spectra can be avoided if adiabatic losses are taken into
account. Another way to keep extremely hard electron distributions in the presence of radiative
losses, is to assume stochastic acceleration models that naturally lead to steady-state relativistic,
Maxwellian-type particle distributions.
We demonstrate that in either case leptonic models can reproduce TeV spectra as hard as
$E_{\gamma}~dN/dE_{\gamma} \propto E_{\gamma}$. Unfortunately this limits, to a large
extend, the potential of extracting EBL from $\gamma$-ray observations of blazars.

\end{abstract}


\section{Introduction}
Blazars constitute a sub-class of Active Galactic Nuclei (AGN) characterized by broadband (from radio to VHE
$\gamma$-rays), non-thermal emission produced in relativistic jets pointing close to the line of sight to the observer
(\citealt{urry95}). The highly variable luminosity of Blazars (which often exhibits 2 peaks) are commonly interpreted
in terms of a synchrotron-inverse Compton origin.

In synchrotron self-Compton (SSC) models, the X-ray emission is usually attributed to the synchrotron radiation of
relativistic electrons. The Compton up-scattering of these synchrotron photons by the same electron population then
 produces the high energy $\gamma$-ray radiation (e.g., \citealt{mara92,bloom96}). Under specific circumstances,
 the target radiation field for inverse Compton upscattering can be dominated by external photons, leading to so-called
 external Compton (EC) models (e.g., \citealt{dermer93,sikora94}). In general, these leptonic models have been relatively
 successful in describing the observed SED of Blazars.

The recent detections of VHE $\gamma$-rays from Blazars with redshift $z \geq 0.1$ (in particular, 1ES 1101-232 at
$z=0.186$ and 1ES 0229+200 at $z=0.139$), however, poses challenges to the conventional leptonic interpretation.
VHE $\gamma$-rays emitted by such distant objects arrive after significant absorption caused by their interactions with
extragalactic background light (EBL) via the process $\gamma\gamma\rightarrow e^+e^-$ (e.g., \citealt{gould67}).
Reconstruction of the absorption-corrected intrinsic VHE $\gamma$-ray spectra based on state-of-the-art EBL models
then yields unusually hard VHE source spectra, that are difficult to account for with standard inverse Compton assumption.

One characteristic case concerns the distant (at $z=0.186$) Blazar 1ES 1101-232, detected at VHE $\gamma$-ray
energies by the H.E.S.S. array of Cherenkov telescopes (\citealt{ahar06,ahar07}). When corrected for absorption
by the EBL, the VHE $\gamma$-ray data result in very hard intrinsic spectra, with a peak in the SED above 3 TeV and
a photon index $\Gamma\leq 1.5$. A similar behavior has also been detected in the TeV Blazar 1ES 0229+200 (at z=0.139)
(\citealt{ahar071ES0229}). Though there is a non-negligible uncertainty in the EBL flux, the intrinsic spectra are unusually
hard even when one considers the lowest levels of the EBL (\citealt{france08}). Other models predicting higher EBL flux
would lead to even harder (intrinsic) photon indices close to 1 (e.g., Stecker \& Scully 2008). We note that a recent analysis
of Fermi LAT data for the nearby TeV Blazar Mkn 501 indicates a hard $\gamma$-ray spectrum ($\Gamma$ close to 1) at
lower (10-200 GeV) energies (\citealt{nero11}). If confirmed, this would be strong evidence for unusually hard $\gamma$-ray
spectra independent of questions related to the level of EBL.
On the other hand and apart from the challenges arising for inverse Compton interpretations, the observed hard VHE spectra
obviously carry important information about the level of the EBL, and thus a deep understanding of the mechanisms acting
within these sources becomes now even more critical.

The "simplest" way to overcome the problem is to assume that there is no absorption. In fact, this is possible in Lorentz
invariance violation scenarios (\citealt{kifu99}). We should note however that this effect is likely to be true only above 2 TeV
(\citealt{stecker01}), whereas the hard spectra problem we face in the case of distant Blazars is relevant to sub-TeV energies
as well. Another non-standard mechanism to avoid severe absorption in the EBL has been suggested by \citealt{deangelis09}
who proposed that $\gamma$-ray photons could oscillate into a new very light axion-like particle close enough to the source and
be converted back before reaching the Earth. To some extend a similar idea was recently suggested by \citealt{essey2011}
who suggested that the $\gamma$-rays from Blazars may be dominated by secondary $\gamma$-rays produced along the
line of sight by the interactions of cosmic rays protons with background photons. While the first scenario would require the
existence of exotic particles, the second needs extraordinary low magnetic fields of the order of $10^{-15} G$.

In more standard-type astrophysical scenarios, the formation of hard $\gamma$-ray spectra is related to the production
and absorption processes. Photon-photon absorption could in principle result in arbitrarily hard spectra provided that the
$\gamma$-rays pass through a hot photon gas with a narrow distribution such that $E_{\gamma}\epsilon_{o}>>m_e c^{2}$.
In this case, due to the reduction of the cross-section, the source becomes optically thick at lower energies and thin at higher
energies, thus leading to formation of hard intrinsic spectra. (\citealt{ahar08,zacha11})

If one relates the hard $\gamma$-ray spectra to the production process, then this implies correspondingly hard parent
particle distributions. Outside standard leptonic models, a number of alternative explanations have been explored in
the literature. In analogy to pulsar winds, \cite{ahar02} for example have analyzed the implications of a cold ultra-relativistic
outflow that initially (close to the black hole) propagates at very high speeds. In such a case, up-scattering of ambient photons
can yield sharp pile-up features in the intrinsic source spectra. However, very high bulk Lorentz factors would be needed
($\Gamma_b \sim10^{7}$) and it seems not clear whether such a scenario can be applied to Blazars. On the other hand,
if Blazar jets would remain highly relativistic out to kpc-scales ($\Gamma_b \sim 10$) and able to accelerate particles, a
hard (slowly variable) VHE emission component could perhaps be produced by Compton up-scattering of CMB photons
(\citealt{bottcher08}).

In order to produce hard $\gamma$-ray spectra within standard leptonic, synchrotron-inverse Compton scenarios, hard
electron energy distributions are required. Although standard shock acceleration theories, both in the non-relativistic
and relativistic regime, predict quite broad, $n(E) \propto E^{-2}$-type, electron energy distributions, there are
non-conventional realizations which could give rise to very hard spectra (\citealt{derishev03,stecker07}).
On a more phenomenological level, \cite{katar07} have shown that the presence of an energetic power-law electron
distribution with a high value of the minimum cut-off energy can lead to a hard TeV spectrum. In general, however,
injection of a hard electron distribution is not a sufficient condition as electrons are expected to quickly lose their energy
due to radiative cooling and thereby develop a standard $n(E) \propto E^{-2}$ form below the initial cut-off energy. In
order to avoid synchrotron cooling, one thus needs to assume unusually small values for the magnetic field within the
source (\citealt{tavecchio09}).

To some extent, this situation can be avoided if one invokes adiabatic losses. This is demonstrated below by means
of a time-dependent investigation assuming dominant adiabatic energy losses. As a second alternative, we discuss
pile-up (Maxwellian-type) electron distributions, that can be formed in stochastic acceleration scenarios. As these
distributions are steady-state solutions with radiative (synchrotron or Thomson) losses already included, there is
no need to avoid these losses. Maxwellian-type electron distributions provide an interesting explanation for the very
hard TeV components as their radiation spectra share many characteristics with the (hardest possible) mono-energetic
distributions.

It is obviously important to explore the strengths and limitations of such explanations in more details, both theoretically
and observationally, in order to understand whether there is a need to invoke more exotic scenarios.

In the present work we explore the conditions under which a narrow, energetic particle distribution is able to successfully
account for the hard VHE source spectra. To this end, we examine different electron distributions within the context of
standard leptonic models, i.e. the one-zone SSC and the external Compton scenario. The paper is structured as following:
The requirements for quasi-stationary SSC solutions are analyzed in Sect.~2. Apart from narrow, power-law-type electron
distributions, quasi-Maxwellian distributions are examined. A time-dependent generalization including adiabatic losses
is explored in Sect.~3. Section~4 discusses the possibilities within an external Compton approach.

\section{Stationary SSC with an energetic electron distribution}
Within a stationary SSC approach, the hardest possible (extended) VHE spectrum is approximately
$F_{\nu} \propto \nu^{1/3}$, where $F_{\nu}=dF/d\nu$ is the spectral flux (differential flux per frequency band).
This has a simple explanation: The emitted synchrotron spectrum of a single electron with Lorentz factor
$\gamma$ in a magnetic field $B$, averaged over the particle's orbit, obeys $j(\nu,\gamma) \propto
G(x)$, where $G(x)$ is a dimensionless function with $x=\nu/\nu_c$ and $\nu_c \equiv 3\gamma^2
e B \sin\alpha/(4\pi m_e c)$. For $x \ll 1$, the functional dependence of $G(x)$ is well approximated
by $G(x) \propto x^{1/3}$, while for $x\gg 1$ one has $G(x) \propto x^{1/2} e^{-x}$ (e.g., \citealt{R&L79}).
Hence, at low frequencies $\nu\ll \nu_c$, the synchrotron spectrum follows $j(\nu) \propto \nu^{1/3}$.
Compton up-scattering of such a photon spectrum in the Thomson regime by a very energetic, narrow
electron distribution will preserve this dependence and therefore yield a VHE spectral wing as hard
as $F_{\nu} \propto \nu^{1/3}$ (see below).

\subsection{Power-law electron distribution with high low-energy cut-off}
A homogeneous SSC scenario with a high value for the low-energy cut-off of the non-thermal electron
distribution has consequently been proposed by \cite{katar07} to overcome the problem of the Klein-Nishina
(KN) suppression of the cross-section at high energies and to reproduce VHE spectra as hard as $1/3$.
Let us assume that the electron population follows a power-law distribution of index $p$ between the
low- and high-energy cut-offs
\begin{equation}\label{el}
N'_{e}(\gamma')=K'_{e}\gamma'^{-p},\;\;\;\;\gamma'_{\rm min}<\gamma'<\gamma'_{\rm max}\,,
\end{equation} as often used in modeling the Blazar spectra. Here, prime quantities refer to the blob rest
frame and unprimed to the observer's frame. Taking relativistic Doppler boosting ($\delta$) into account,
the observed synchrotron flux from an optically thin source at distance $d_L$ is given by the integral of
$N'_e(\gamma') d\gamma'$ times the single particle emissivity $j'(\nu',\gamma')$ over the volume element
and all energies $\gamma'$ (e.g., \citealt{begelman84}), i.e
\begin{equation}
 F^{\rm syn}_{\nu}=\frac{\delta^{3}}{d^{2}_{L}}\int_{V'}\int_{\gamma'}j'(\nu',\gamma')N'_e(\gamma') d\gamma' dV'\,.
\end{equation}
The above expression yields the common power-law of index $\alpha=(p-1)/2$ between the frequency limits
$\nu_{\rm min}\propto \delta (B\gamma^{2}_{\rm min})$ and $\nu_{\rm max}\propto \delta (B \gamma^{2}_{\rm max})$.
Below and above those limits, the electrons with energy around the minimum and maximum cut-off dominate,
and thus the spectrum approximately exhibits a slope $F_{\nu}\propto\nu^{1/3}$ for $\nu<\nu_{\rm min}$, and
an exponential cut-off for $\nu>\nu_{max}$, i.e.
 \begin{equation}
F_{\nu}\propto\left\{
  \begin{array}{lll}
  \nu^{1/3},\;\;\;\nu<<\nu_{min} \\
   \nu^{-\frac{p-1}{2}},\;\;\;\nu_{min}\leq\nu\leq\nu_{max}\\
  \nu^{1/2}e^{-\nu},\;\;\;\nu>>\nu_{max} \\
  \end{array}
     \right.
\end{equation}

The hard $1/3$-slope appears in the VHE range of EC $\gamma$-rays when the synchrotron photons are
up-scattered to higher energies by the electron population given by equation~(\ref{el}) with a high
$\gamma_{\rm min}$ and provided that the Thomson regime applies. Obviously, it will be significantly softer
in the KN regime. In any case, however, there exists a characteristic energy below which the Compton
spectrum mimics the behavior of the synchrotron spectrum $F_{\nu}\propto\nu^{1/3}$.

Note that the inverse Compton-scattered spectrum of a monochromatic photon field by mono-energetic
electrons approximately follows, at low up-scattered photon energies, $F_{\nu}\propto\nu$ (cf. \citealt{B&G70}).
Thus, any photon field which is softer (flatter) than $F_{\nu}\propto\nu$ will dominate the lower-energy
part of the up-scattered emission and thus, in the standard SSC scenario the $1/3$-VHE slope (the
$4/3$-slope in the $\nu F_\nu$ representation) is the hardest that can be achieved.

An exemption to this may occur if the magnetic field in the source would be fully turbulent with zero mean
component. In such a case, the low-frequency part of the synchrotron spectrum could be harder than
$F_{\nu} \propto \nu^{1/3}$ (\citealt{medvedev06,derishev07,revi10}), which will then be reflected to
low-energy part of the Compton component.

\begin{figure}
\epsscale{0.3pt}
\includegraphics[width=360pt]{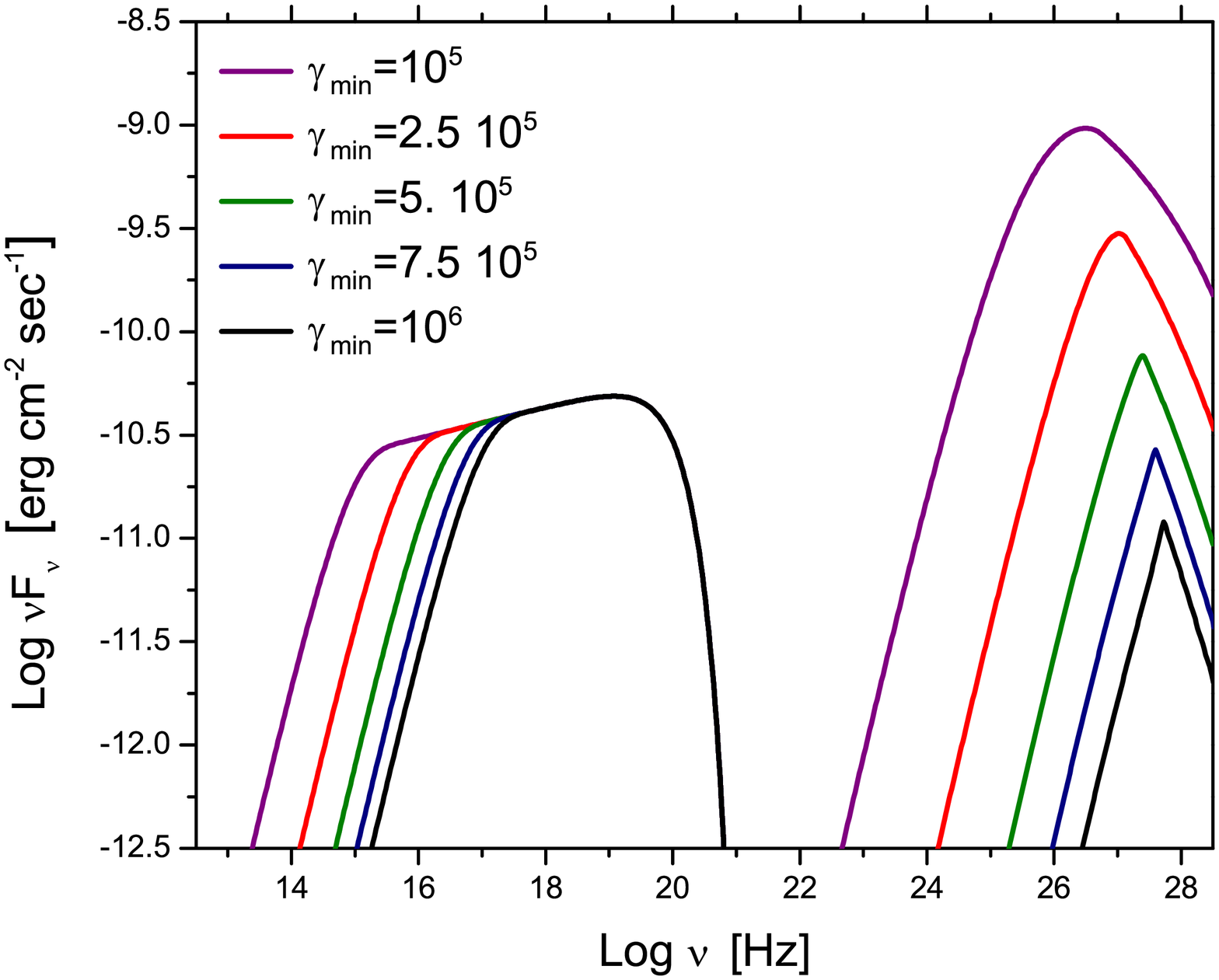}
\caption{Stationary SSC spectra for different values of the low-energy cut-off. Above $\gamma_{\rm min}\sim 3 \times10^{5}$ we are very deep in the Klein-Nishina regime and the peak of the Compton emission appears very sharp. As one reduces $\gamma_{\rm min}$, the suppression of the cross-section decreases and the minimum Compton energy drops to lower energies. Thus, the peak of the Compton flux raises significantly, whereas the synchrotron peak remains constant. A Doppler factor $\delta=50$ has been used
for the plot.}\label{different_gmin}
\end{figure}

The "critical Compton energy" is usually $\epsilon_{\rm min} \simeq \delta \gamma_{\rm min}^2 (b
\gamma^{2}_{\rm min})$, where $b\equiv(B/B_{cr}) m_e c^{2}$, $B_{cr}=m_e^2c^3/(e\hbar)$, except when
the deep KN regime applies, i.e., when up-scattering of the minimum synchrotron photons by the minimum
energy electrons occurs in the KN regime so that $\frac{4}{3}b \gamma^{3}_{\rm min}>1$. If the latter applies,
then the corresponding energy below which one can see the hard $1/3$-slope is, as expected,
$\gamma_{\rm min}m_e c^{2}$, and it approximately corresponds to the peak of the emitted luminosity for
any power-law electron index (see Fig. \ref{different_gmin}).
In the KN regime, the peak appears especially sharp (e.g., \citealt{tavecchio98}), and the Compton flux
has a strong inverse dependence on the value of $\gamma_{\rm min}$. For example, for the realization
presented in Fig.~(\ref{different_gmin}), the emissivity in this regime roughly scales as $j^C \propto
\gamma_{\rm min}^{-2.5}$, so that slight changes in $\gamma_{\rm min}$ can lead to significant variations
in the amplitude of the Compton peak flux. On the other hand, as long as $p<3$ (positive synchrotron slope
in a $\nu F_{\nu}$ representation) the synchrotron peak luminosity would remain approximately constant.

\begin{figure}
\epsscale{0.3pt}
\includegraphics[width=360pt]{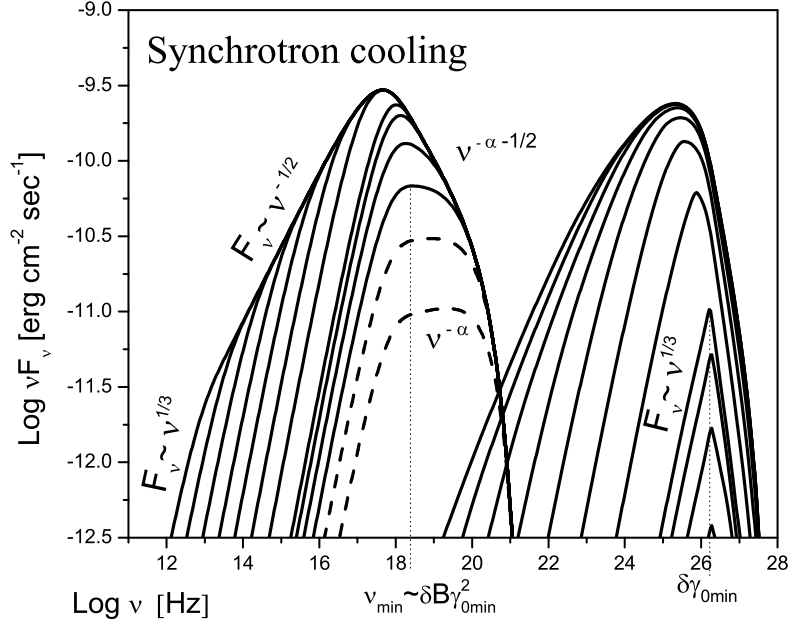}
\caption{Evolution of the observed SSC spectrum for constant injection of a narrow power-law electron distribution, with modifications due
to synchrotron cooling taken into account. The magnetic field is $B=1G$. The hard (1/3) synchrotron and Compton spectral wings are
observed for timescales shorter than the cooling timescale of the $\gamma_0$-particles, i.e., in the present application for timescales $\leq
0.1$ days. The figure shows the expected spectral evolution for a total (observed) time $t \sim 1$ day. Parameters used are $R_0 =7.5\times
10^{14}$ cm, $\gamma_{\rm min}=7\times10^{4}$, $\gamma_{\rm max}=2\times 10^{6}$, power law index $p=2.85$ and Doppler factor
$\delta=25$. The total injected power is $Q\sim 10^{41}$ erg/sec.}\label{SYNlosses}
\label{syncooling}
\end{figure}

A power-law electron distribution with a high low-energy cutoff has been used in \cite{tavecchio09} in order
to reproduce the SED of the blazar 1ES~0229+200 within a stationary SSC approach. The high value of
$\gamma_{\rm min} \sim 10^{5}$ then ensures the hard Compton part of the spectrum with 1/3-slope is in
the TeV range. The generic difficulty for such an approach is that an energetic electron distribution is expected
to quickly develop a $\gamma^{-2}$-tail below $\gamma_{\rm min}$ due to synchrotron cooling, thereby
making the Compton VHE spectrum softer (see Fig.~\ref{SYNlosses}). To overcome this problem,
\cite{tavecchio09} suggested an unusually low value for the magnetic field, $B\sim (10^{-4}-10^{-3})$ G,
that would allow the electron distribution to remain essentially unchanged on timescales of up to a few
years. Obviously, one would then not expect to observe significant variability on shorter timescales. We
note however, that this requirement could be relaxed if one assumes that the detected $\gamma$-ray
signal is a superposition of short flares which can not be detected individually. Arguments based on
magnetic flux conservation naively suggest that the magnetic field value, when scaled from the black hole
region to the emission site, should be at least one or two orders of magnitude higher so that one would
need to destroy magnetic flux for such a scenario to work. On the other hand, a narrow but very energetic
electron distribution in combination with such low magnetic field strengths implies a strong deviation from
equipartition, thereby obviously facilitating an expansion of the source.

\subsection{Relativistic Maxwellian electron distribution}
As far as a narrow energetic particle distribution is concerned, a relativistic Maxwellian may come as a more
natural representation. Such an electron distribution can be the outcome of a stochastic acceleration process
(e.g., 2nd order Fermi) that is balanced by synchrotron (and/or Compton) energy losses, or in general any
energy loss mechanism that exhibits a quadratic dependence on the particle energy (see e.g.,
\citealt{Schlickeiser85}; \citealt{aharonian86}; \citealt{henri91}; \citealt{lukas2008}).

Consider for illustration the Fokker-Planck diffusion equation which describes the stationary distribution
function $f(p)$ of electrons that are being accelerated by, e.g., scattering off randomly moving Alfv\'{e}n
waves in an isotropic turbulent medium,
\begin{equation}\label{dif}
\frac{1}{p^{2}}\frac{\pad}{\pad p}\left(p^{2}D_{p}\frac{\pad f(p)}{\pad p}\right)+\frac{1}{p^{2}}
\frac{\pad}{\pad p}\left(\beta_{s}p^{4}f(p)\right)=0\,,
\end{equation}
where $D_p$ is the momentum-space diffusion coefficient. Particle escape is neglected in eq.~(\ref{dif}),
as the timescale for synchrotron cooling is expected to be much smaller than the one for electron escape.

For scattering off Alfv\'{e}n waves, one has $D_p=\frac{p^{2}}{3\tau}(\frac{V_{A}}{c})^{2}\equiv D_0
p^{2-\alpha_p}$, with $V_{A}=\frac{B}{\sqrt{4\pi \rho}}$ the Alfv\'{e}n speed and $\tau =\lambda/c\propto
p^{\alpha_p}$, $\alpha_p \geq 0$, the mean scattering time (e.g., \citealt{rieger07}). If the turbulent wave
spectrum $W(k) \propto k^{-q}$ is assumed to be Kolmogorov-type ($q\simeq5/3$) or Kraichnan-type($q=
3/2$), the momentum-dependence becomes $\alpha_p = 1/3$ and $\alpha_p=1/2$, respectively. Bohm-type
diffusion, on the other hand, would imply $\alpha_p=1$, while hard-sphere scattering is described by
$\alpha_p=0$. Note, however, that if one considers electron acceleration by resonant Langmuir waves,
even $D_p =$ const ($\alpha_p=2$) may become possible \citep{aharonian86}.

The synchrotron energy losses that appear in the second term of Eq.~(\ref{dif}) are
\begin{equation}
\frac{dp}{dt}=-\beta_{s}
p^{2}=-\frac{4}{3}(\sigma_{T}/m_e^2c^2) U_{B}p^{2}
\end{equation}

In the $\gamma$-parameter space, the solution of
Eq.~(\ref{dif}) becomes a relativistic Maxwell-like function
\begin{equation}\label{max}
f(\gamma)=A\gamma^{2}e^{-\left(\frac{\gamma}{\gamma_{c}}\right)^{1+\alpha_p}}\,,
\end{equation}
($\alpha_p \neq -1$) with
\begin{equation}
\gamma_c=\left(\frac{[1+\alpha_p] D_0}{\beta_s}\right)^{1/(1+\alpha_p)} (m_e c^2)^{-1}\,,
\end{equation}
and constant $A$ to be defined by the initial conditions. Note that this is a steady-state solution already
including radiative losses and there is no need to invoke extreme values for the magnetic field. The critical
Lorentz factor $\gamma_c$ approximately corresponds to the energy at which acceleration on timescale
\begin{equation}
t_{\rm acc}=\frac{3}{4-\alpha_p}\left(\frac{c}{V_{A}}\right)^{2} \tau
\end{equation} is balanced by (synchrotron) cooling on timescale $t_{\rm cool}=1/[\beta_s p]$. Depending
on the choice of parameters, a relatively large range of values for $\gamma_{c}$ is possible and thus, cut-off
energies of the order of $\gamma_{c}\sim 10^{5}$ may well be achieved. Consider, for example, Bohm-type
diffusion with $\tau =\eta r_g/c$, $r_g=\gamma m_e c^2/(e B)$ the electron gyro-radius and $\eta \geq 1$.
Using $t_{\rm acc} =t_{\rm cool}$, the maximum electron Lorentz factor becomes $\gamma_c \simeq
10^6~(v_A / 0.01 c)~(1~\mathrm{G}/B)^{1/2}\eta^{-1/2}$.

\begin{figure}
\epsscale{0.3pt}
\includegraphics[width=360pt]{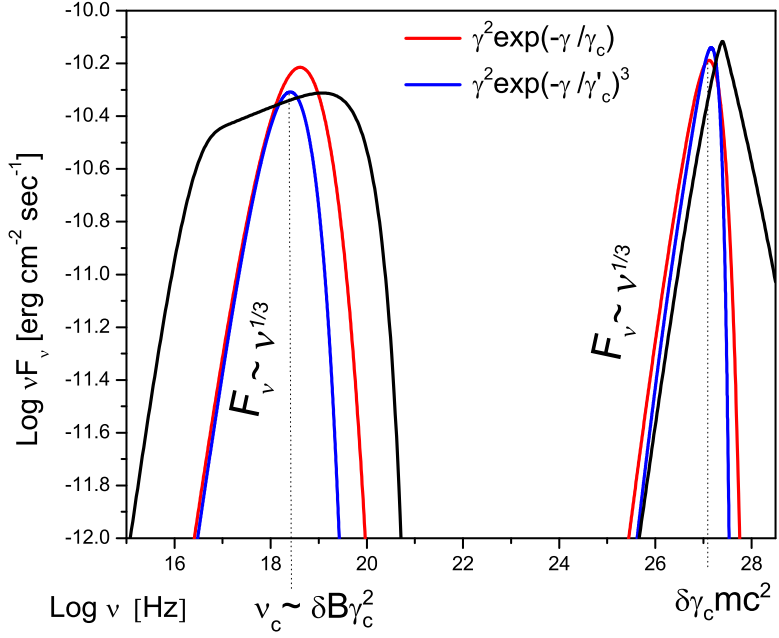}
\caption{SSC modelling with different electron distributions. \textbf{Black line:} {\it Power-law} with large value of the minimum energy (as in
Tavecchio et al.~2009). The parameters used are $\gamma_{\rm min}=5\times10^{5}$, $\gamma_{\rm max} = 4 \times 10^{7}$, power law
index $p=2.85$, $B=4\times10^{-4}$ G, $k_{e}=6.7\times 10^{8}$ cm$^{-3}$, $R=5.4\times10^{16}$ cm and Doppler factor $\delta=50$.
\textbf{Red line:} Relativistic {\it Maxwellian} distribution $N_{e}=K_e\gamma^{2}\exp(-\frac{\gamma}{\gamma_{c}})$ with parameters $\gamma_{c}=1.5\times10^{5}$, $B=0.07$ G, $K_{e}=3\times10^{-14}$cm$^{-3}$, $R=2\times10^{14}$cm and
$\delta=33$. The peak of Compton flux occurs in the KN regime as $(B/B_{cr})\gamma^{3}_{c}\simeq160>>1$. 
\textbf{Blue line:} Relativistic {\it Maxwellian} distribution $N_{e}=K_e\gamma^{2}\exp(-\frac{\gamma}{\gamma_{c}})^{3}$
with parameters $\gamma_{c}=5.3\times 10^{5}$, $B=0.06$ G, $K_{e}=4\times 10^{-15}$ cm$^{-3}$, $R=2 \times 10^{14}$ cm
and $\delta=33$.}\label{SSCmax}
\end{figure}

The synchrotron spectrum that arises from a Maxwell-like electron distribution is dominated by the emission
of electrons with $\gamma_{c}$ (Fig. \ref{SSCmax}). It exhibits the characteristic $1/3$-slope up to the
corresponding "synchrotron cut-off frequency" $h \nu^{\rm syn}_{c}\sim\delta b\gamma^{2}_{c}$ where
$b=B/B_{cr}$ and $B_{cr}=m^2c^3/e\hbar$. Thus the Compton spectrum is very similar to the one resulting
from a narrow power-law if one chooses a value for the cut-off energy close to the minimum electron energy
of the power-law distribution. The peak of the Compton flux then contains information for the cut-off energy
as $\nu^{c}_{\rm peak}\propto \gamma_{c}$.

Note that for an electron distribution of the form of eq.~(\ref{max}) that exhibits an exponential cutoff $\propto
\exp[-(\gamma/\gamma_{c})^{\beta}]$, the corresponding cut-off in the synchrotron spectrum appears much
smoother, $\propto \exp[-(\nu/\nu_{*})^{\beta/(\beta+2)}]$ (\citealt{frit89,zirak07}). The position of the synchrotron
peak flux, $\nu_p$, is then also dependent on $\beta$, and one can show that for $\beta=1$ (or $\alpha_p=0$ in
the previous notation) an important factor $\sim 10$ arises, so that $\nu_{p}=9.5\nu_{c}$, whereas for $\beta=3$
the synchrotron peak corresponds approximately to the electron cut-off as $\nu_{p}=1.2\nu_{c}$ (e.g., Fig.
\ref{SSCmax}).

\section{Time-dependent case - expansion of the source}
Expansion of the source could change the conclusions drawn above. In particular, if one assumes a very low
magnetic field such that synchrotron losses are negligible, then adiabatic losses may become important and
alter the electron distribution. In this section, we examine the behavior of the system for a power-law electron
distribution with a high value of the low-energy cut-off discussed above. For simplicity, we consider a spherical
source that expands with a constant velocity $u$,
\begin{equation}
R(t)=R_0+u(t-t_0)\,.
\end{equation}
The relativistic electron population will be affected by synchrotron losses,
\begin{equation}
P_{\rm syn}=-\frac{d\gamma}{dt}=\frac{\sigma_{T}B(t)^{2}\gamma^{2}}{6\pi m_e c}\,,
\end{equation}
and by adiabatic losses (e.g., Longair 1982),
\begin{equation}\label{adiabatic}
P_{\rm ad}=-\frac{d\gamma}{dt}\simeq \frac{1}{3}\frac{\dot{V}}{V}\gamma=\frac{\dot{R}(t)}{R(t)}~\gamma
                       =\frac{u}{R(t)}~\gamma\,.
\end{equation}
\noindent As the emission region expands, the magnetic field decreases. We consider a scaling $B \propto (1/R)^m
\propto (1/t)^m$ with $1\leq m\leq 2$ to study the evolution of the system. The limiting value $m=2$ corresponds to
conservation of magnetic flux for the longitudinal component, whereas $m=1$ holds for the perpendicular component.
(Note that for $m=1$ the ratio of the electrons' energy density to the magnetic field energy density remains constant).
Which energy loss process then determines the electron behavior depends mainly on the magnetic field strength and
the size of the source. A simple comparison of the above relations shows that when $P_{\rm ad}>P_{\rm syn}$, i.e.,
\begin{equation}
B(t)^{2}R(t)<\frac{6 \pi m_e c^{2}}{\sigma_{T}}\left(\frac{u}{c}\right) \frac{1}{\gamma}=2.3 \times 10^{19}
                        \left(\frac{u}{c}\right)\frac{1}{\gamma}
\end{equation}
adiabatic losses dominate over radiative losses. For example, if one considers expansion at speed $u \sim c$ and an
initial source dimension $R_0\sim 10^{14}$ cm, then for energies below $\gamma \sim 10^{7}$ the magnetic field can
be as large as $B\sim 0.1$ G and for energies less than $\gamma\sim 10^{5}$ the adiabatic losses are still dominant
for a value of $B\sim 1$ G. If the expansion of the source would not affect the hard slope at TeV energies, this could
thus allow for a relaxation of the values used for SSC modeling of the source. In order to investigate this scenario in
more detail, one needs to solve the electrons' kinetic equation
\begin{equation}
\frac{\pad N_{e}(\gamma,t)}{\pad t}=\frac{\pad}{\pad\gamma}\left(P_{\rm ad}N_{e}(\gamma,t)\right)-\frac{N_{e}(\gamma,t)}{\tau_{e}}
                          +Q(\gamma,t)\,,
\end{equation}
where $\tau_{e}$ is the characteristic escape time and $N_{e}$ the differential electron number. For simplicity, we
neglect the escape term ($\tau_{e}\rightarrow \infty$), assuming that the sources expands with relativistic speeds
$u\sim 0.1$c. For a constant expansion rate and continuous injection with rate $Q(\gamma,t) \rightarrow Q(\gamma, R)$,
we can replace the time variable $t$ by the source dimension $R$. Then, the general solution of the kinetic equation
(eq. (30) in \citealt{atoyan99}) for the case of dominance of adiabatic losses is reduced to
\begin{equation}
N_e(\gamma,R)=\frac{R}{R_0}~N_0\left(\frac{R}{R_0}\gamma\right)+
\frac{1}{u}\int^{R}_{R_0}\frac{R}{r}~Q\left(\frac{R}{r}\gamma,r\right)dr\,,
\end{equation}
where the first term corresponds to the initial conditions, the contributions of which quickly disappears, and the
second term relates to the continuous injection of relativistic electrons. $R_0$ is the source dimension at the
initial time $t_0$.

We consider zero initial conditions ($N_0=0$) and power-law injection of relativistic particles at constant rate
\begin{equation}\label{injection}
Q(\gamma,R)=Q_0\gamma^{-p_1}~\Theta(\gamma-\gamma_{0,\rm min})\Theta(\gamma_{0,\rm max}-\gamma)
\Theta(R-R_0)\,,
\end{equation}
where $\Theta$ denotes the unit step function
\begin{equation}\label{theta}
\Theta(x-x_{0})=\left\{
\begin{array}{ll}
1,\;\;\;\;x>x_{0} \\
0,\;\;\;\;x<x_{0}\,, \\
\end{array}
\right.
\end{equation}
$p_1>0$ is the momentum index and $R_0$ the radius at which injection starts. At radius $R$, electrons
with initial cut-off energies $\gamma_{0, \rm max}$ and $\gamma_{0, \rm max}$ will have energies
$\gamma_{R, \rm min}$ and $\gamma_{R, \rm max}$, respectively, as they evolve according to
Eq.~(\ref{adiabatic}), i.e. we have
\begin{equation}
\gamma_{R}=\gamma_0 \frac{R_0}{R(t)}\propto \frac{1}{t}\,.
\end{equation}
Moreover, there exists a critical radius $R_{*}=R_0\gamma_{0, \rm max}/\gamma_{0, \rm min}$ at which
$\gamma_{R, \rm max}$, i.e. the energy of the electron with initial injected energy $\gamma_{0,\rm max}$ at
$R$, becomes less than the initial $\gamma_{0,\rm min}$, so that the following two cases can be distinguished:\\

\noindent For $R<R_{*}$, or equivalently as long as $\gamma_{\rm R,max}>\gamma_{\rm 0,min}$, we have
\begin{equation}\label{sol1}
N(\gamma,R)=\frac{Q_0}{p_1 u}R\left\{
\begin{array}{lll}
\gamma^{-p_1}\left[1-\left(\frac{\gamma}{\gamma_{\rm 0,max}}\right)^{p_1}\right],\;\;\;\;\gamma_{\rm R,max}<\gamma<\gamma_{\rm 0,max} \\
\: & \: & \: \\
\gamma^{-p_1}\left[1-\left(\frac{R_0}{R}\right)^{p_1}\right],\;\;\;\;\gamma_{\rm 0,min}<\gamma<\gamma_{\rm R,max} \\
\: & \: & \: \\
\gamma_{\rm 0,min}^{-p_1}-\left(\frac{R \gamma}{R_0}\right)^{-p_1},\;\;\;\;\gamma_{\rm R,min}<\gamma<\gamma_{\rm 0,min} \\
\end{array}
\right.
\end{equation}

\noindent For $R>R_{*}$, or equivalently as long as $\gamma_{\rm R,max}<\gamma_{\rm 0,min}$, the solution is
\begin{equation}\label{sol1}
N(\gamma,R)=\frac{Q_0}{p_1 u}R\left\{
\begin{array}{lll}
\gamma^{-p_1}\left[1-\left(\frac{\gamma}{\gamma_{0,max}}\right)^{p_{1}}\right],\;\;\;\;\gamma_{0,min}<\gamma<\gamma_{\rm 0,max} \\
\: & \: & \: \\
\gamma^{-p_1}_{\rm 0,min}-\gamma^{-p_1}_{\rm 0,max},\;\;\;\;\gamma_{\rm R,max}<\gamma<\gamma_{\rm 0,min} \\
\: & \: & \: \\
\gamma_{\rm 0, min}^{-p_1}-\left(\frac{R\gamma}{R_0}\right)^{-p_1},\;\;\;\;\gamma_{\rm R, min}<\gamma<\gamma_{\rm R, max} \\
\end{array}
\right.\\
\end{equation}

\begin{figure}
\epsscale{0.3pt}
\includegraphics[width=360pt]{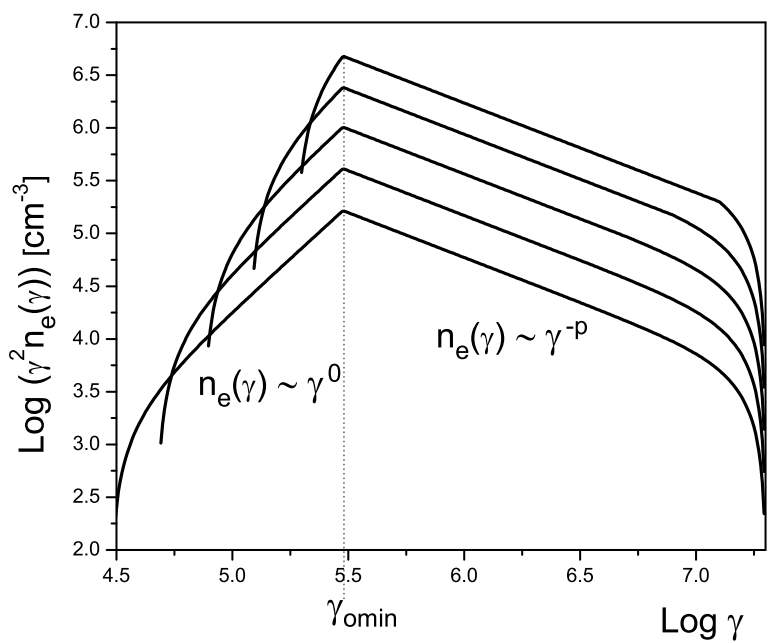}
\caption{Illustration of the evolution of the electron distribution for constant injection and dominant adiabatic losses. The
expansion of the source does not modify the power-law index above the initial low-energy cut-off $\gamma_{\rm 0, min}$,
whereas below it the distribution becomes approximately flat. The electron number density depends on radius as $n_{e}
\propto R^{-2}$}
\label{electrons}
\end{figure}

\begin{figure}
\epsscale{0.3pt}
\includegraphics[width=360pt]{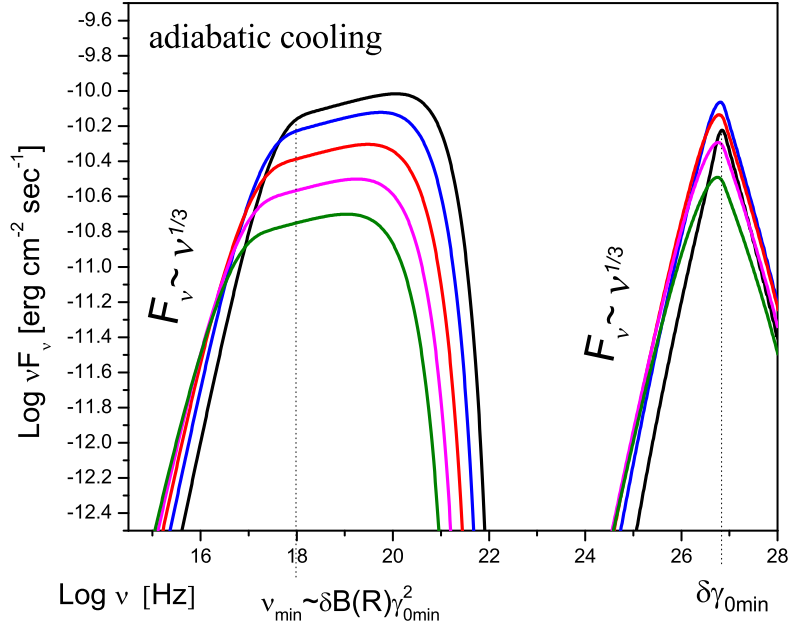}
\caption{Evolution of the observed SSC spectrum with constant injection of a narrow power-law and dominant adiabatic losses. As the
source evolves, the synchrotron peak decreases and gets shifted to smaller energies following the decrease of the magnetic
field. For the magnetic field, we use an initial value $B_0=0.075$ G and we assume that it scales as $B=B_0~(R_0/R)$ (i.e., $m=1$).
The initial radius is $R_0 =7.5\times 10^{14}$ cm, expanding up to $R=10 R_0$ (at $u=0.1$ c) and corresponding to observed
timescales of the order of $t\sim 30/\delta$ days. The total injected power is $Q \sim 5\times 10^{41}$ erg/sec. Other parameters
used are $\gamma_{\rm min}=3\times10^{5}$, $\gamma_{\rm max}=2 \times 10^{7}$, power law index $p_1=2.85$, Doppler factor $\delta=25$
and $Q_0=1.5\times10^{52}$ sec$^{-1}$. Note that timescales are comparable to the synchrotron cooling case.}
\label{ADBL1}
\end{figure}

\begin{figure}
\epsscale{0.3pt}
\includegraphics[width=360pt]{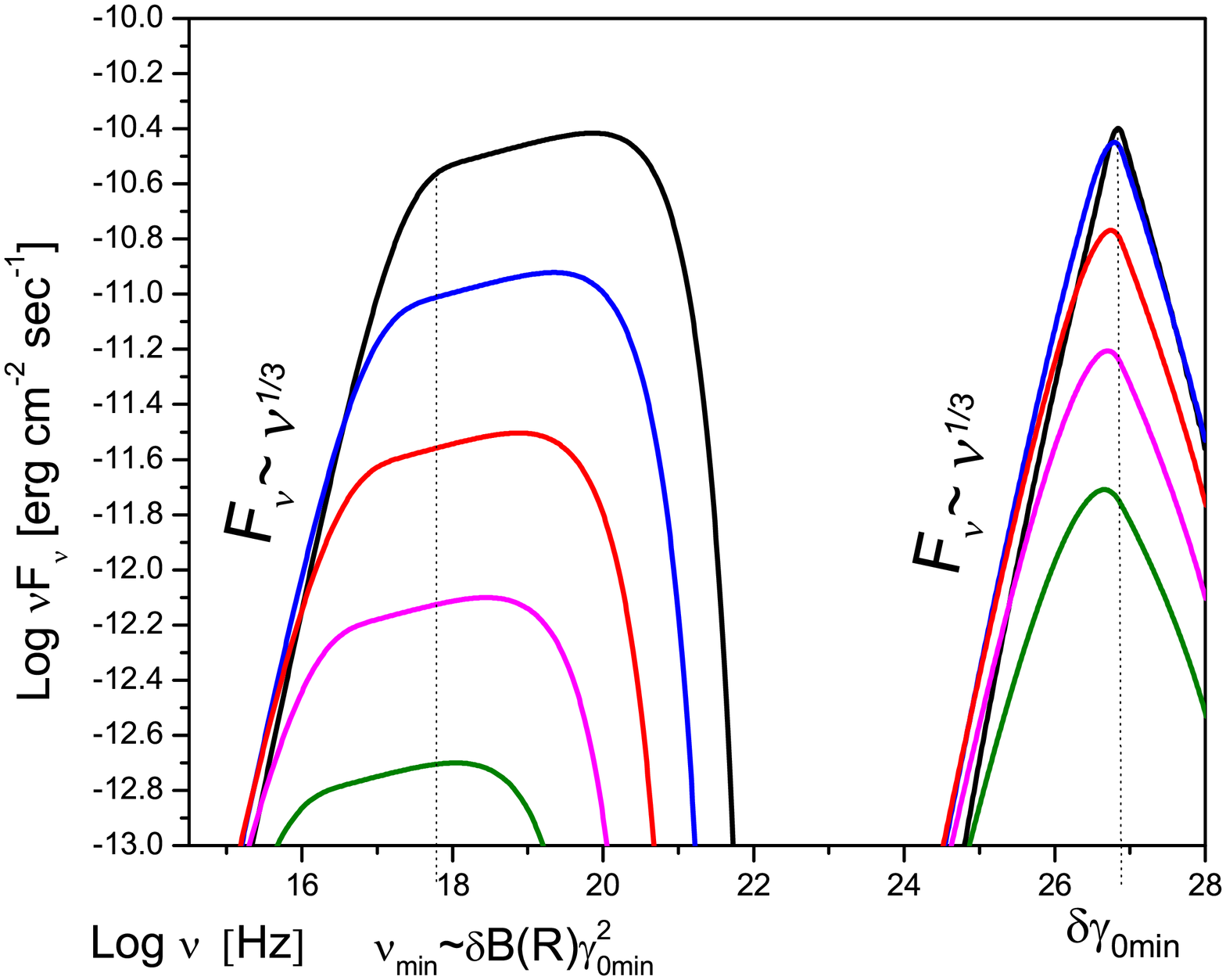}
\caption{Same as Fig.~\ref{ADBL1} but for a different magnetic field scaling, $m=2$. Other parameters are the same as in Fig.~\ref{ADBL1}.} \label{ADBL2}
\end{figure}

\begin{figure}
\epsscale{0.3pt}
\includegraphics[width=360pt]{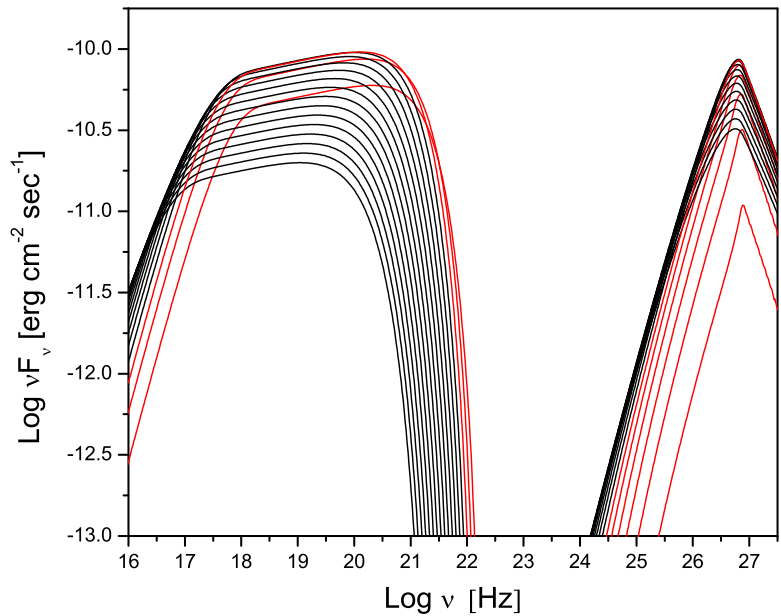}
\caption{Evolution of the SSC spectrum with dominant adiabatic losses (for $B \propto 1/R$, i.e., $m=1$) from $R_0 = 7.5 \times 10^{14}$ cm
to $R=10 R_0$, and corresponding observed luminosities. Whereas the synchrotron peak gets with time significantly shifted to lower energies,
the  Compton peak can appear almost static. The Compton flux reaches its maximum at a greater radius than for the synchrotron one. Synchrotron
and Compton fluxes are shown until maximum (with red lines) and after (with black lines).}
\label{peak_evolution}
\end{figure}

\begin{figure}
\epsscale{0.3pt}
\includegraphics[width=360pt]{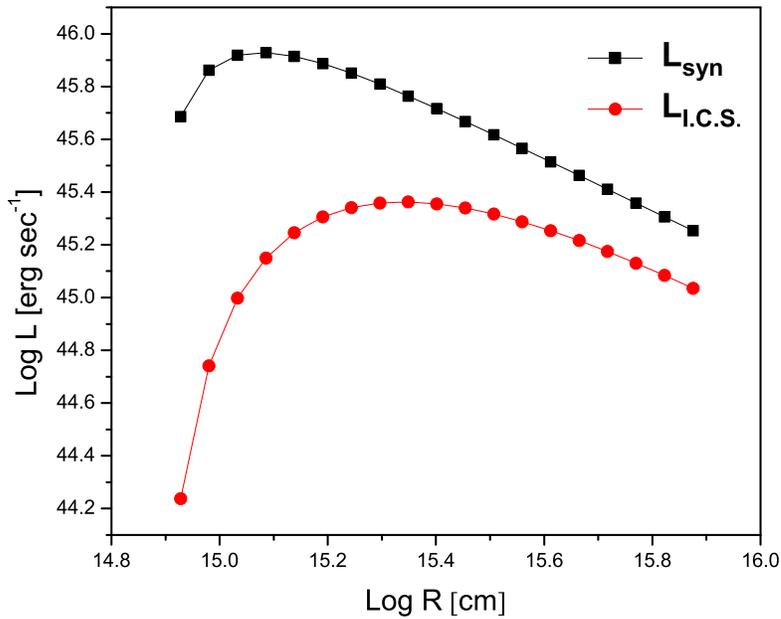}
\caption{Evolution of the SSC spectrum luminosities with dominant adiabatic losses (for  $m=1$), see also Fig.~\ref{peak_evolution}.
The Compton flux reaches its maximum at greater radius (i.e., later) compared to the synchrotron one. While during the raising
phase the variability pattern approximately shows a quadratic behavior, the correlation becomes almost linear during
the declining phase.}
\label{peak_flux_evolution}
\end{figure}

\noindent The two solutions exhibit the same behavior. The differential electron number density drops with radius
as $n_{e}(\gamma,R)=\frac{N_{e}(\gamma,R)}{\rm Volume}\propto R^{-2}$, and above the initial low-energy cut-off
$\gamma_{\rm 0,min}$ adiabatic losses do not modify the power-law index $n_{e}(\gamma,t)\propto
\gamma^{-p_1}$ (\citealt{kardashev62}). Below $\gamma_{\rm 0,min}$ the resulting distribution is constant with
respect to the electron energies, $n_{e}(\gamma,R)\propto\gamma^{0}$ (Fig.~\ref{electrons}). The $\gamma^{0}$-part
of the electron population does not show up in the spectrum as the contribution of the injected $\gamma_{\rm 0,min}$
electrons (generating a 1/3-synchrotron wing) remains dominant at  low energies (Figs.~\ref{ADBL1} and \ref{ADBL2}).
Thus, in contrast to the simple synchrotron cooling case, one has $F_{\nu}\propto \nu^{1/3}$ below the injected cut-off.
For this reason, the classical hard spectrum picture at the TeV range can remain for timescales analogous to the source
size. Even though electrons cool adiabatically as the source expands, the hard $1/3$-synchrotron slope always appears
below the synchrotron frequency related to the initial minimum Lorentz factor
\begin{equation}
\nu^{\rm syn}_{\rm min}\propto \gamma^{2}_{\rm 0,min} B(R)\propto \frac{1}{t^m}\,.
\end{equation}
Note that any decrease of this break energy occurs due to a decrease of the magnetic field. This is different to the pure
synchrotron cooling case, where the corresponding break energy follows the evolution of the minimum electron energy
so that $\nu^{\rm syn}_{\rm min}\propto 1/t^{2}$. The same consideration holds for the energy regime where Compton
scattering occurs. When we are deep in the KN regime ($b(R)\gamma^{3}_{\rm 0,min}>1$) the energy below which the
hard slope remains is, as mentioned above,
\begin{equation}
\nu^{C}_{\rm min}\propto\gamma_{\rm 0, min}\,.
\end{equation}
It therefore does not move to lower energies though the corresponding synchrotron frequency does. (In the pure
synchrotron cooling case, obviously, $\nu^{C}_{\rm min}\propto 1/t^{2}$). As the source expands and the magnetic field
drops, there will be an instant $t$ corresponding to a radius $R$ at which the KN regime no longer applies, and the
break Compton frequency becomes
\begin{equation}
\nu^{C}_{\rm min}\propto B(R)\gamma^{4}_{\rm 0, min}\propto\frac{1}{t^m}\,,
\end{equation}
which now moves to lower frequencies with the same rate as the synchrotron one. (Note that this reveals a very
different time-dependence compared to the pure synchrotron cooling case where now $\nu^{C}_{\rm min}\propto
1/t^{4}$). However, the peak of the Compton flux still remains close to the initial $\gamma_{\rm 0, min}$ energy
(Fig.~\ref{peak_evolution}), and in total the decrease of the synchrotron peak flux is much stronger than the decrease
in the Compton peak flux (Fig.~\ref{peak_flux_evolution}). In general, the dependence of the magnetic field on the
radius $R$ has important consequences for the behavior of the system even though the synchrotron losses are not
important. The synchrotron peak flux varies as
\begin{equation}
F^{syn}_{\nu}\propto N_{e} B(R)^{2}
\end{equation}
and as we know from the solution of the kinetic equation that $N_e\propto R$, the variability of the synchrotron
luminosity should reflect the magnetic field dependence. The Compton flux, on the other hand, does not necessarily
vary quadratically with respect to the synchrotron flux. As discussed above, the drop of the minimum Compton energy
(which occurs naturally within the expansion-scenario) reduces the suppression of the cross-section and thereby
supports the Compton emission. The variability pattern after "saturation" can therefore approach a quasi-linear dependence
(cf. Fig~\ref{peak_flux_evolution}). Initially, during the raising phase before the two luminosities reach their maximum,
the Compton flux can vary much more strongly, almost more than quadratically, with respect to the synchrotron flux.
Moreover, close to saturation the Compton luminosity can exhibit a delay with respect to the synchrotron one as it
reaches its maximum at later times compared to the synchrotron luminosity.

The above considerations apply to situations where the expansion of the source completely determines the evolution
of the system. In reality, synchrotron losses could modify the electron distribution at high energies, namely for
\begin{equation}
\gamma>\gamma_{*}=\frac{6\pi m_e c^{2}}{\sigma_T R B(R)^2}\left(\frac{u}{c}\right)\,.
\end{equation}
However, as synchrotron losses decrease faster than adiabatic losses, one only needs to ensure that initially $\gamma_*
>\gamma_{\rm 0, min}$. The change of the electron power index from $-p_1$ to $-p_1-1$ due to synchrotron cooling
(cooling break) above $\gamma_{\rm 0, min}$ would then not disturb the hard $1/3$ slope in the TeV range.

\section{The external Compton case}\label{external}
An alternative hypothesis to the SSC scenario concerns the Comptonization of a radiation field external to the electron
source. In general, the optical-UV radiation field produced by a standard accretion disk could represent a non-negligible
external source of photons to be up-scattered to the VHE $\gamma$-ray part of the spectrum. This radiation field could
be up-scattered either directly by the relativistic electrons of the jet (with target photons coming directly from the accretion
disk, \citealt{dermer93}) or more effectively after being reprocessed/re-scattered by emission line clouds like the broad
line region (BLR) (\citealt{sikora94}). In external Compton (EC) scenarios, the geometry of the source and the location
of the photon field with respect to the jet are of high importance as they can result in strong boosting or de-boosting effects
on the photon energies. Here we explore the possibility of producing a hard TeV spectrum within the EC approach.
We consider the BLR case, where the photon field is strongly boosted in the frame of the jet and up-scattered to higher
energies.

Let us consider a blob of relativistic electrons that travels with the jet of bulk Lorentz factor $\Gamma$ along the $z$-axis.
The jet passes through a region assumed to be filled with isotropic and homogeneous photons that obey a Planckian
distribution of temperature $T\sim 10^{4}-10^{5}K$ (corresponding peak frequency $\nu_d = 2.82 kT/h \sim 5\times
10^{14}-10^{15}$ Hz). The central disk photon field is then characterized by a special intensity
\begin{equation}
I^{d}_{\nu}=\frac{2h\nu^{3}}{c^{2}(e^{\frac{h\nu}{kT}}-1)}\,,
\end{equation}
a fraction $\xi <1$ of which we assume is isotropized by re-scattering or reprocessing in the BLR (\citealt{sikora2002}),
so that the spectral energy density of the target photon field is
\begin{equation}\label{externalfield}
U^{\rm BLR}_{\nu}=\frac{\xi L^{d}_{\nu}}{4\pi cr^{2}_{\rm BLR}}\,,
\end{equation}
where $L^{d}_{\nu}=4\pi^2 r^2_{d} I^d_{\nu}$ is the spectral luminosity of the disk and where we take $r_d \sim r_s$ for
the disk radius.

In order to take anisotropic effects into account, we transform the electron distribution from the comoving blob frame $K'$
to the rest frame $K$ of the external photon field, which in our case coincides with the observer's frame (\citealt{georga01}).
Electrons are assumed to be isotropic in the blob frame $K'$. In the photon frame $K$ they exhibit a strong dependence
on the angle $\theta$, which is the observer's angle. As the up-scattered photons travel inside a cone $1/\gamma$, we can make the approximation that they follow the direction of the electrons. The angle between the electron momentum and
the bulk velocity of the jet coincides with the observer's angle. The observer practically sees radiation only from electrons
that in the photon frame are directed towards him. The observed flux then is
\begin{equation}\label{ECflux}
F_{\epsilon_{\gamma}}=\frac{\delta^{3}}{d_{L}^{2}}\int N'_{e}(\frac{E}{\delta})~W(E,\epsilon_{\rm ph},\epsilon_{\gamma})~
                      n_{\rm ph}(\epsilon_{\rm ph})dE~d\epsilon_{\rm ph}\,
\end{equation}
where $N_{e}(E)$ denotes the differential number of electrons per energy per solid angle, and $W=E_{\gamma}
\frac{dN}{dtdE_{\gamma}}$ is the scattered photon spectrum per electron (given in \citealt{B&G70}). The unprimed
quantities refer to the external photon field frame with number density $n_{\rm ph}(\epsilon_{\rm ph})$ and the primed
ones to the blob rest frame.

\begin{figure}
\epsscale{0.3pt}
\includegraphics[width=360pt]{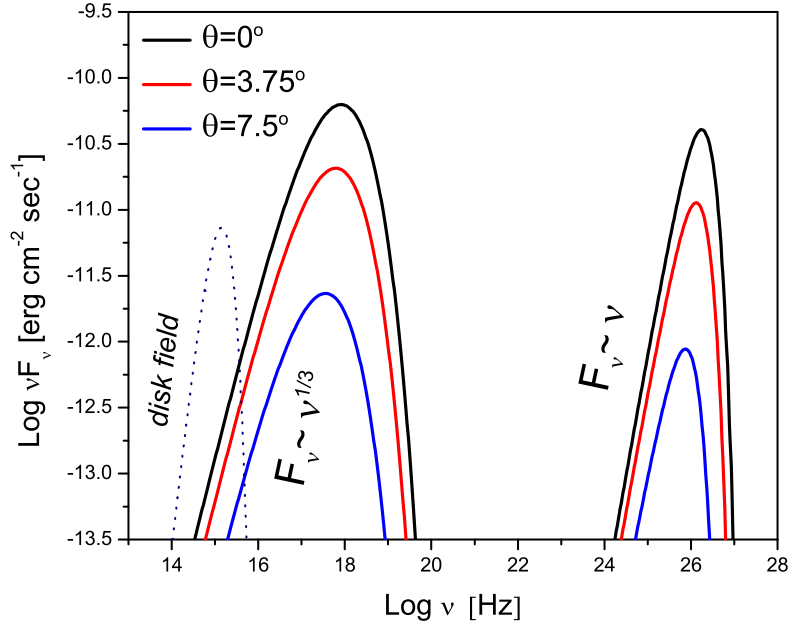}
\caption{External Compton scenario for a Maxwellian-type electron distribution (with $\alpha_p=0$). The observed spectrum is calculated for
different angles $\theta$ to the observer. The synchrotron slope follows $F_{\nu}\propto\nu^{1/3}$. In the TeV range $F_{\nu}\propto\nu^{1}$, i.e.,
harder than in the SSC case. The dashed line corresponds to the assumed disk spectrum. The bulk Lorentz factor of the jet is $\Gamma=13$ and
the peak energy of the electron distribution is $\gamma_{c}=2\times 10^{4}$. For the disk photon field a temperature $T=1.75\times 10^{4}$
K is assumed. The relevant radius $R_d$ of the disk is considered to be of the same dimension as the jet ($10^{15}$ cm). The magnetic
field is $B=1$ G and a fraction $\xi=0.1$ of the disk photons is assumed to be rescattered by the BLR.}\label{testEC}
\end{figure}

We show the calculated Compton spectrum for a Maxwellian electron distribution in Fig.~\ref{testEC}. The resulting
TeV slope appears even harder than in the SSC case, with a limiting value of $F_{\nu}\propto\nu^{1}$. Any photon
field which is softer (flatter) than $F_{\nu}\propto \nu^{1}$ will dominate the Compton spectrum at low energies, as in
our SSC model case where the up-scattered (synchrotron) photon spectrum follows $F_{\nu}\propto\nu^{1/3}$. In all
other cases, like in the external Compton scenario with a Planckian photon field (that at low energies follows
$F_{\nu}\propto \nu^{2}$), the characteristic behavior of the Compton cross-section appears, implying that the Compton
spectrum at low energies (i.e., below $\sim \gamma_c^2 \epsilon_c$, where $\gamma_c$ is the electron break
frequency in the photon rest frame) is dominated by the contribution from the up-scattering of the peak photons with
$\epsilon_{c}\sim 3 kT$, yielding a $F_{\nu} \propto \nu$ dependence.

A similar consideration holds for a narrow (energetic) power-law electron distribution in an expanding source scenario.
A hard VHE component $F_{\nu}\propto \nu$ should then appear below $\delta \gamma_{\rm 0,min}$. The critical energy
below which one can see this hard behavior of the Compton flux will not move to lower energies as the external target
photon field is quasi-stable, so that the condition for the deep Klein-Nishina regime $\gamma_{\rm 0,min}\epsilon_{c}>1$
does not change.

\section{Summary}
 The observed hard $\gamma$-ray spectra of TeV Blazars are difficult to explain within the most popular leptonic
 synchrotron-Compton models. The $n(E) \propto E^{-2}$ shape, that the electron energy distribution is expected
 to quickly develop due to synchrotron cooling, usually results in a $F_{\nu}\propto\nu^{-1/2}$ radiation spectrum
 and therefore represents the limiting value of how hard the up-scattered spectrum can be in the TeV range. Moreover,
 modification due to Klein-Nishina effects can make the up-scattered TeV spectrum steeper than this and shift the
 Compton peak to much lower energies than observed.

However, intrinsic source spectra as hard as $F_{\nu}\propto \nu^{-1/2}$ exist, even when one only corrects for the
lowest level of the EBL, most notably in the case of 1ES~1101-232 and 1ES~0229+200 (\citealt{ahar06, ahar071ES0229}).
Most likely, the real source spectra are even harder. Investigating the possibility of forming hard VHE blazar spectra
appears therefore particularly important. Methodologically, it seems necessary to first examine the "conventional"
radiation and acceleration mechanisms, that have often been successful in interpreting Blazar observations, before
adopting very different and often more extreme solutions.

The results of this work show that within a simple homogeneous one-zone SSC approach, a power-law particle
distribution with a large low-energy cutoff can in principle produce a hard ($\alpha=1/3$) -- slope in the VHE
domain ($F_\nu\propto \nu^{\alpha}$) by reflecting the characteristic low-energy slope of the single particle
synchrotron spectrum (cf. also \citealt{katar07}). As shown in section \ref{external}, even harder VHE spectra
approaching $F_{\nu}\propto\nu$ ($\alpha=1$) can be achieved in the external Compton case for a Planckian-type
ambient photon field.

A power-law electron distribution with a high low-energy cut-off has been used in \cite{tavecchio09} to model the emission
from 1ES 0229+200 within a stationary SSC approach. In order to avoid the above noted synchrotron cooling problem, an
unusually low value for the magnetic field strength was employed, leaving the particle distribution essentially unchanged
on the timescales of several years. This goes along with a strong deviation from simple equipartition by several orders of
magnitude (i.e., $u_B/u_e \simlt 10^{-5}$). While it is known from detailed spectral and temporal SSC studies of the prominent
$\gamma$-ray Blazar Mkn~501 that TeV sources may be out of equipartition at the one per-mill level or less (\citealt{kraw02}),
the SSC modeling of 1ES 0229+200 suggest the hard spectrum sources to belong to the more extreme end. (Within external
Compton models values closer to equipartition may be achieved, depending on the external photon field energy density).
On the other hand, a large electron energy density (strongly exceeding the magnetic field one) could well facilitate an
expansion of the source, and this motivates a time-dependent analysis:

Using a time-dependent SSC model, we have shown that the hard ($\alpha=1/3$)-VHE slope can be recovered, when
adiabatic losses dominate over the synchrotron losses for the low-energy part of the electron distribution (i.e., for Lorentz
factors less than the injected $\gamma_{\rm min}$). The main reason for this is, that the resultant electron distribution
below $\gamma_{\rm min}$ becomes flat and therefore does not show up in the SSC spectrum. Interestingly, this scheme
also allows one to relax the very low magnetic field constraints.

We also examined the relevance of a Maxwellian-like electron distribution that peaks at high electron Lorentz factors
$\sim 10^5$. Such a distribution represents a simple time-dependent solution that already takes radiative energy losses
into account, and turns out to be capable of successfully reproducing the hard spectra in the TeV range (with limiting values
$\alpha=1/3$ and $\alpha=1$, respectively). Maxwellian distributions can be the outcome of a stochastic acceleration
process balanced by synchrotron or Thomson cooling. Depending on the physical conditions within a source, e.g., if
particles undergo additional cooling in an area different from the acceleration one (\citealt{sauge06,giebels07}), or if
the medium is clumpy supporting a "multi-blob" scenario in which the observed radiation is the result of superposition
of regions characterized by different parameters, the combination of pile-up distributions may allow a suitable interpretation
of different type of sources. For the case presented here, they demonstrate a physical way of achieving the high low-energy
cut-offs needed in leptonic synchrotron-Compton models for the hard spectrum sources.

\begin{figure}
\epsscale{0.1pt}
\includegraphics[width=300pt,angle=-90]{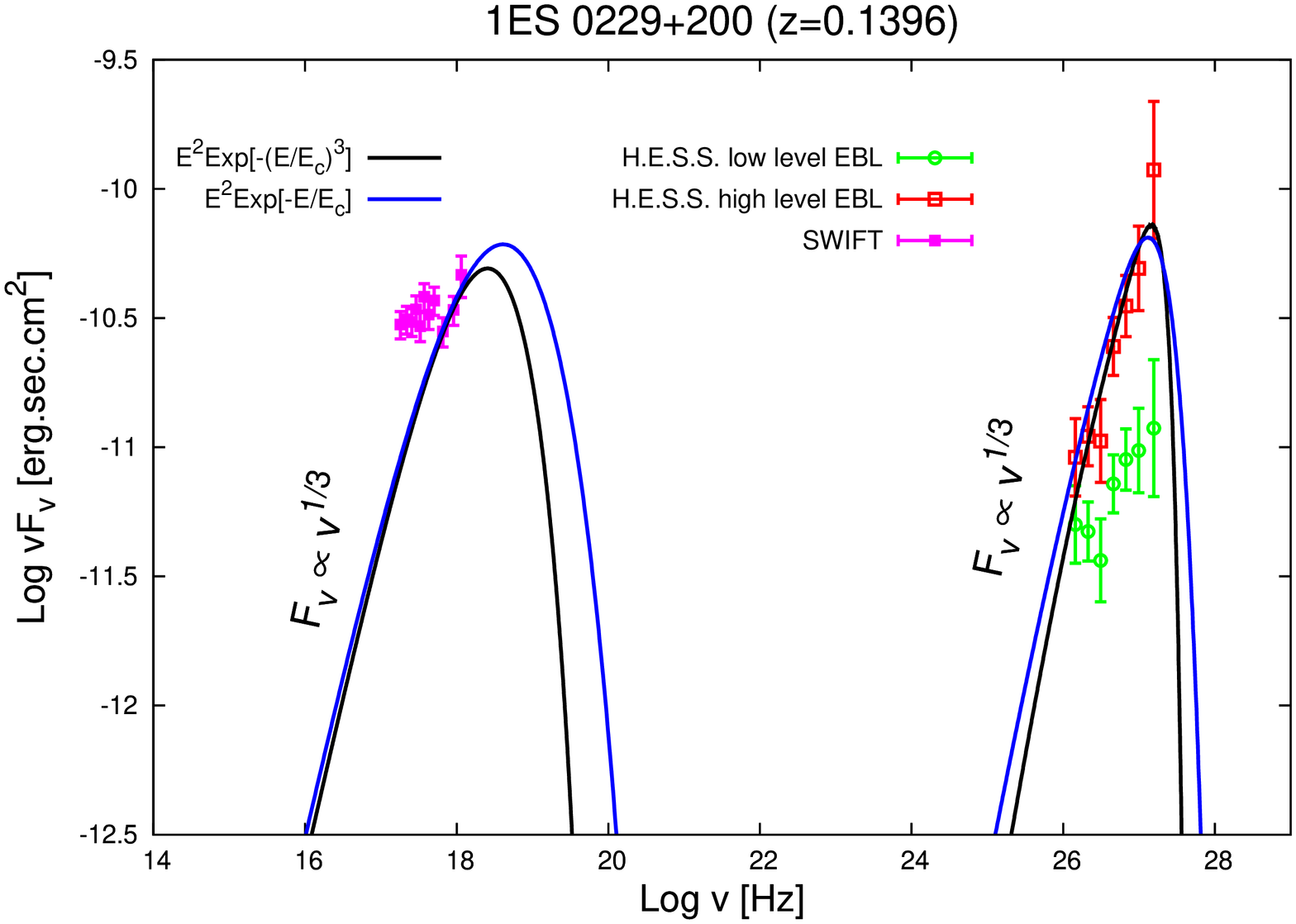}
\caption{The hard spectrum blazar 1ES~0229+200 at z=0.139 with SED modeled within an SSC approach using Maxwellian-type electron
distributions. All parameters used are the same as in Fig.\ref{SSCmax}. Data points shown in the figure are from \cite{zacha11}, where the
intrinsic (de-absorbed) source spectrum has been derived based on the EBL model of \cite{france08} with (i) EBL level as in their original
paper ("low level EBL") and (ii) (maximum) EBL level scaled up by a factor of 1.6 ("high level EBL").}\label{apply}
\end{figure}

Although our main purpose here is not to fit data, Fig.~\ref{apply} shows that a Maxwellian-type electron distribution
could also provide a satisfactory explanation for the hard TeV component in 1ES 0229+200.

Our results illustrate that even within a leptonic synchrotron-Compton approach relatively hard intrinsic TeV source
spectra may be encountered under a variety of conditions. While this may be reassuring, the possibility of having
such hard source spectra within "standard models" unfortunately constrain the potential of extracting limits on the EBL
density based on $\gamma$-ray observations of Blazars, one of the hot topics currently discussed in the context of next
generation VHE instruments.\\

{\it Acknowledgement: We would like to thank S. Kelner and S. Wagner for helpful discussions.}

{}

\begin{thebibliography}{}

\bibitem[\protect\citeauthoryear{Aharonian et al.}{1986}]{aharonian86}
Aharonian, F.A., Atoyan, A. M., $\&$ Nahapetian, A. 1986, A$\&$A, 162, L1

\bibitem[\protect\citeauthoryear{Aharonian et al.}{2002}]{ahar02}
Aharonian, F.A., Timokhin, A.N., \& Plyasheshnikov, A.V. 2002, A\&A, 384, 834

\bibitem[\protect\citeauthoryear{Aharonian et al.}{2006}]{ahar06}
Aharonian, F., et al. 2006, Nature, 440, 1081

\bibitem[\protect\citeauthoryear{Aharonian et al.}{2007a}]{ahar07}
Aharonian, F., et al., 2007a, A$\&$A, 470, 475

\bibitem[\protect\citeauthoryear{Aharonian et al.}{2007b}]{ahar071ES0229}
Aharonian, F., et al.,  2007b, A\&A, 475, L9

\bibitem[\protect\citeauthoryear{Aharonian et al.}{2008}]{ahar08}
Aharonian, F.A., Khangulyan, D., $\&$ Costamante, L. 2008, MNRAS, 387, 1206

\bibitem[\protect\citeauthoryear{Atoyan $\&$ Aharonian}{1999}]{atoyan99}
Atoyan A. M., $\&$ Aharonian F. A. 1999, MNRAS, 302, 253

\bibitem[\protect\citeauthoryear{Begelman et al.}{1984}]{begelman84}
Begelman, M.C., Blandford, R.D., \& Rees, M.J. 1984, Reviews of Modern Physics, 56, 255

\bibitem[\protect\citeauthoryear{Bloom $\&$ Marcher}{1996}]{bloom96}
Bloom, S.D. \& Marscher A.P. 1996, ApJ, 461, 657

\bibitem[\protect\citeauthoryear{Blumenthal $\&$ Gould}{1970}]{B&G70}
Blumenthal, G.R. $\&$ Gould, R.J. 1970, Reviews of Modern Physics, 42, 237


\bibitem[\protect\citeauthoryear{B\"ottcher et al.}{2008}]{bottcher08}
B\"ottcher, M., Dermer, C.D., $\&$ Finke, J.D. 2008, ApJ, 679, L9

\bibitem[\protect\citeauthoryear{De Angelis et al.}{2009}]{deangelis09}
De Angelis, A., Mansutti, O., Persic, M., \& Roncadelli, M. 2009, MNRAS, 394, L21

\bibitem[\protect\citeauthoryear{Derishev}{2007}]{derishev07}
Derishev, E. V. 2007, Ap$\&$SS, 309, 157

\bibitem[\protect\citeauthoryear{Derishev et al.}{2003}]{derishev03}
Derishev, E.V., Aharonian, F.A., Kocharovsky, V.V., Kocharovsky, Vl. V. 2003, Phys.
Rev. D, 68, 043003

\bibitem[\protect\citeauthoryear{Dermer et al.}{1993}]{dermer93}
Dermer C.D., \& Schlickeiser R. 1993, ApJ, 416, 458


\bibitem[\protect\citeauthoryear{Essey et al.}{2011}]{essey2011}
Essey, W., Kalashev, O., Kusenko, A., $\&$ Beacom, J. F. 2011 ApJ, 731 51E

\bibitem[\protect\citeauthoryear{Franceschini et al.}{2008}]{france08}
Franceschini, A., Rodigliero, G., $\&$ Vaccari, M. 2008, A$\&$A, 487, 837

\bibitem[\protect\citeauthoryear{Fritz}{1989}]{frit89}
Fritz, K.D.\ 1989, A\&A, 214, 14

\bibitem[\protect\citeauthoryear{Georganopoulos et al.}{2001}]{georga01}
Georganopoulos, M., Kirk, J.G., $\&$ Mastichiadis, A. 2001, ApJ, 561, 111

\bibitem[\protect\citeauthoryear{Giebels et al.}{2007}]{giebels07}
Giebels, B., Dubus, G., \& Kh\'{e}lifi, B. 2007, A\&A, 462, 29

\bibitem[\protect\citeauthoryear{Gould $\&$ Schr{\'e}der}{1967}]{gould67}
Gould, R.J., $\&$ Schreder, G.P. 1967, Physical Review, 155, 1408

\bibitem[\protect\citeauthoryear{Henri $\&$ Pelletier}{1991}]{henri91}
Henri, G., \& Pelletier, G.\ 1991, ApJL, 383, L7

\bibitem[\protect\citeauthoryear{Kardashev}{1962}]{kardashev62}
Kardashev, N. S. 1962, Soviet Astronomy, 6, 317

\bibitem[\protect\citeauthoryear{Katarzy{\'n}ski et al.}{2007}]{katar07}
Katarzy{\'n}ski, K., Ghisellini, G., Tavecchio, F., Gracia, J., $\&$ Maraschi, L. 2006, MNRAS, 368, L52

\bibitem[\protect\citeauthoryear{Kifune}{1999}]{kifu99}
Kifune, T.\ 1999, ApJL, 518, L21

\bibitem[\protect\citeauthoryear{Krawczynski et al.}{2002}]{kraw02}
Krawczynski, H., Coppi, C.S., \& Aharonian, F. 2002, MNRAS, 336, 721

\bibitem[\protect\citeauthoryear{Maraschi et al.}{1992}]{mara92}
Maraschi, L., Ghisellini, G., \& Celotti, A.\ 1992, ApJL, 397, L5

\bibitem[\protect\citeauthoryear{Medvedev}{2006}]{medvedev06}
Medvedev, M. V. 2006, ApJ, 637, 869

\bibitem[\protect\citeauthoryear{Neronov et al.}{2011}]{nero11}
Neronov, A., Semikoz, D. \& A.M. Taylor 2011, A\&A submitted (arXiv:1104.2801)

\bibitem[\protect\citeauthoryear{Reville $\&$ Kirk}{2010}]{revi10}
Reville, B., $\&$ Kirk, J.  G.\ 2010, ApJ, 724, 1283

\bibitem[\protect\citeauthoryear{Rieger et al.}{2007}]{rieger07}
Rieger, F.M., Bosch-Ramon, V., $\&$ Duffy, P. 2007, Ap$\&$SS, 309, 119

\bibitem[\protect\citeauthoryear{Rybicki $\&$ Lightman}{1979}]{R&L79}
Rybicki, G.B., $\&$ Lightman, A.P. 1979, Radiative Processes in Astrophysics, Wiley, New York

\bibitem[\protect\citeauthoryear{Saug\'{e} \& Henri}{2006}]{sauge06}
Saug\'{e}, L., \&  Henri, G. 2006, A$\&$A, 454, L1

\bibitem[\protect\citeauthoryear{Schlickeiser}{1985}]{Schlickeiser85}
Schlickeiser, R. 1985, A$\&$A, 143, 431

\bibitem[\protect\citeauthoryear{Sikora et al.}{1994}]{sikora94}
Sikora, M., Begelman, M. C., $\&$ Rees, M. J. 1994, ApJ, 421, 153

\bibitem[\protect\citeauthoryear{Sikora et al.}{2002}]{sikora2002}
Sikora, M., Blazejowski, M., Moderski, R., $\&$ Madejski, G. M. 2002, ApJ, 577, 78

\bibitem[\protect\citeauthoryear{Stawarz \& Petrosian}{2008}]{lukas2008}
Stawarz, L., $\&$ Petrosian, V. 2008, ApJ, 681, 1725

\bibitem[\protect\citeauthoryear{Stecker et al.}{2007}]{stecker07}
Stecker, F.W., Baring, M. G. $\&$ Summerlin, E. J. 2007, ApJL, 667, L29.

\bibitem[\protect\citeauthoryear{Stecker \& Scully}{2008}]{stecker08}
Stecker, F.W., $\&$ Scully, S.T. 2008, A \&A, 478, L1.

\bibitem[\protect\citeauthoryear{Stecker \& Glashow}{2001}]{stecker01}
Stecker, F.W., $\&$ Glashow, S.L. 2001, Astropart. Phys. 16, 97

\bibitem[\protect\citeauthoryear{Tavecchio et al.}{1998}]{tavecchio98}
Tavecchio, F., Maraschi, L., $\&$ Ghisellini, G. 1998, ApJ, 509, 608

\bibitem[\protect\citeauthoryear{Tavecchio et al.}{2009}]{tavecchio09}
Tavecchio, F., Ghisellini, G., Ghirlanda, G., Costamante, L., $\&$ Franceschini, A. 2009, MNRAS, 399, L59

\bibitem[\protect\citeauthoryear{Urry \& Padovani}{1995}]{urry95}
Urry, C.M., \& Padovani, P.\ 1995, PASP, 107, 803

\bibitem[\protect\citeauthoryear{Zacharopoulou et al.}{2011}]{zacha11}
Zacharopoulou, O., Khangulyan, D., Aharonian, F., $\&$ Costamante, L. 2011, ApJ in press

\bibitem[\protect\citeauthoryear{Zirakashvili \& Aharonian}{2007}]{zirak07}
Zirakashvili, V.N., \& Aharonian, F.\ 2007, A\&A, 465, 695


\end{thebibliography}
\end{document}